# Ideal Glasses and Protein Network Dynamics

J. C. Phillips,

Dept. of Physics and Astronomy, Rutgers University, Piscataway, N. J., 08854-8019

ABSTRACT

Quantitative topological analogies between the flexibilities of optimized inorganic glasses, small biological molecules, and proteins suggest that mean field estimates of internal stress are useful in identifying mechanisms supporting living materials and their enzymatic products. These analogies bode well for emerging minimalist flexibility models of protein dynamics. Application to trehalose, the optimal bioprotective material, leads to a remarkably simple mechanical model, closely parallel to mechanical effects observed in sandwich-like proteins.

Molecular networks occur commonly in inorganic glasses (such as window glass), while proteins are globularly compacted peptide chains with multiple network contacts of side groups and hydrogen bonds. Thus the picture of protein dynamics as characterized by many of the same features that are universally present in network glasses, polymers and hydrogen-bonded molecular liquids (such as glycerol) is gaining in popularity [1-3].

Glasses are very long-lived metastable phases that exhibit universal features (such as stretched exponential relaxation (SER)) not observed in solids or liquids in equilibrium. It is tempting to suppose that these universal features are describable by axiomatic mean-field theories, while the variations specific to individual materials involve local field corrections. Mean-field and local-field effects are lumped together in molecular dynamics simulations (MDS), making it difficult to separate them. The growing power and accuracy (using realistic interatomic force fields, often designed to reproduce the results of first-principle quantum calculations) of MDS is reflected in their rapidly increasing number (thousands/year). At the same time, realistic (large scale) MDS of hydrogen-bonded materials are often time-limited (to maximum times of order 10 ps or perhaps ns), whereas protein dynamics functions on time scales of ms or s. Mean field properties can often persist over these much longer times, so their identification and separation can amplify the content and significance of otherwise time-limited MDS.

The program of separating and quantifying mean- and local-field effects is already well advanced in both experiment and theory of inorganic network glasses, especially many chalcogenide alloy glasses [4]. The critical variable is the connectivity of the network, which can be under-, over-, or optimally constrained [5]. In mean-field models the transition from under- to over-constrained occurs abruptly at an ideal composition $x_c$, while when local field effects are added this transition broadens and occurs over a composition range $x_1 < x < x_2$. The $[x_1, x_2]$ window usually includes $x_c$, and it has remarkable properties: within the window the density is a maximal mesa (not a peak), the glass transition is nearly reversible, and aging virtually ceases [4]. Moreover, these highly favorable properties appear to be related directly to the internal stress of the network, which nearly vanishes in the window [6].

The time limitations of MDS effectively prevent predictions of the composition dependence of the glass-forming tendency in network glasses. Given the obvious hierarchy of constraining interactions [bond-stretching, -bending, dihedral angles,…], and the obvious glass-forming condition ($N_c$ = number of constraints, $N_d$ = number of degrees of freedom = 3(number of atoms $N_A$), all per formula unit) that $N_c = N_d$, mean-field theory initially successfully predicted [7] the composition dependence of the glass-forming tendency in network glasses, thus identifying the ideal compositions $x_c$ in binary alloys ($As_2Se_3$, $GeSe_4$, $SiO_2$, and $(Na_2O)(SiO_2)_4$). The global (nonlocal) condition of vanishing network stress is one of the two conditions satisfied by window glass (the other is optimized space-filling local topology, corresponding to maximal H bonding here) [8].

There is nothing in mean-field network theory [5,7- 9] that limits its applications to inorganic networks, so here it is applied two important properties of H-bonded networks: the protein – water interface, and the glass-forming tendency in H-bonded molecular liquids, such as glycerol. The rules for calculating network constraints are straightforward, and have been discussed extensively in papers on inorganic glasses [9]; for the reader's convenience they are summarized and illustrated in an Appendix to this paper [10]. The new rules needed here for H bonding are that each C-H--- bond always adds one stretching constraint, and each O-H--- two constraints, one stretching and one bending. Depending on morphology and $T_g$, there may also be additional C-H--- bond bending constraints. One cautionary remark: while it might appear that counting constraints is the same as counting spring forces in a vibrational model, this is not always the case. Specifically there are $N(N-1)/2 = 6$ three-body bond-bending springs for an $N = 4$-fold coordinated atom, but only 5 *linearly independent* bond-bending constraints.

The highly specific power of the zero stress or $N_c = N_d$ condition is illustrated by the fact that most simple hydrides are not glass formers at all, because they are severely underconstrained, even when the maximum number of H bonding interactions (which may actually be sterically suppressed, as for methane, Table I) is assumed. A familiar example of an overconstrained network is a-Si, prepared not by melt quenching, but by evaporation at low temperatures. Heating films of a-Si does not lead to a glass transition, but to explosively exothermic crystallization. Most polymers and copolymers (such as polybutadiene, $(CH_2)_n$) are overconstrained when H-bond interactions are included, which is why these materials often are partially crystallized, even when crystallization is heterogeneously suppressed by $H_2O$ inclusions. In general the ideal glass-forming condition of nearly zero internal network stress is rarely realized.

Thermal expansion at constant pressure ($\alpha_P = V^{-1}(\partial V/\partial T)_P$) and at constant dielectric relaxation time ($\alpha_\tau = V^{-1}(\partial V/\partial T)_\tau$) near the glass transition temperature ($\tau = 1$ s) provide an easy way [11] to gauge quantitatively whether or not H bonding is critically enhancing the glass-forming tendency. (The results are quantitatively similar for $\tau = 1000$ s, or when $\tau$ is replaced by the viscosity $\eta$ [12].) The measured values of the ratio $\alpha_\tau/\alpha_P$ are near unity (they range from 0.6 to 2.8) for 15 molecular and polymer glass-formers, but are 6 and 17 for the strongly hydrogen bonded polyalcohols sorbitol and glycerol, respectively. In Table I the intramolecular covalent constraints $N_{cc}$ are counted separately from the intermolecular H-bonding constraints $N_{cH}$. We see that glycerol is an ideal glass, with $N_{cc} + N_{cH} = 42 = 3(14) = N_d$, while sorbitol is slightly underconstrained, $N_{cc} + N_{cH} = 75$, while $N_d = 78$, corresponding to a reduction in $\alpha_\tau/\alpha_P$ by a factor of 3. These examples illustrate not only the accuracy of constraint counting, but also its simplicity: $\alpha_\tau$ in particular is not directly calculable by methods presently available.

While this result is in gratifyingly good agreement with the chemical trends in $\alpha_\tau/\alpha_P$, it is inconclusive: glycerol, for example, is much shorter than sorbitol, so a larger pressure dependence is to be expected for the latter. Instead, a truly exacting test of the H-bond constraint count near $T_g$ is provided by a 45-atom molecule, trehalose, which commonly exists as a dihydrate. Trehalose has unique properties: it is a nonreducing glycoside that has been found in large quantities in organisms (algae, bacteria, fungi, insects, invertebrates, and yeasts as well as a few flowering plants) that are able to survive extreme external stresses such as high or very low temperatures or periods of complete drought up to 120 years (anhydrobiosis). These qualities led to the suggestion [13], now apparently prescient, that trehalose forms a glassy structure around embedded biomolecules and inhibits thus the denaturization due to formation of ice crystals. However, many still wonder if the intrinsic properties of the vitreous state itself constitute a crucial element that plays a key role in the bioprotective action of trehalose [14]. Recent neutron scattering studies [15] showed stretched exponential relaxation [$\exp[-(t/\tau)^\beta]$] with a dimensionless stretching fraction $\beta = 0.38$ in good agreement with the values near 0.4 observed in many other glasses influenced by both short-range and long-range forces [16]. In Table I we see that for trehalose, $N_{cc} = 98$, $N_{cOH} = 18$, and $N_d = 3(45) = 135 = N_{cc} + N_{cH}$, so that trehalose is also an ideal glass like glycerol. Figs. 1 and 2 of [17] and here show both molecules; their morphologies are quite different, but both form ideal glasses when maximally hydrogen bonded.

Once we know that trehalose is an ideal glass former, it is easy to understand its bioprotective functionality. Under extreme conditions, water may either evaporate or freeze, in both cases with disastrous consequences. Dessication or cooling stiffens the protein network, but ideal glasses are rigid yet stress free [6]. They therefore support the normal morphology, and resist fracture in general terms. The larger the molecule, the more effective its stabilizing interaction. This is why trehalose is more effective in evolutionary morphologies, although the difference in its bioprotective properties from other carbohydrates (such as sucrose) is often small in synthetic contexts.

Trehalose has an elegant structure composed of two sugar flaps connected by a bridging O. The interactions of the two flaps around their dihedral angle are weak, giving the flaps the ability to adapt retentively to the substrate that they are protecting. In calculating the intermolecular bending constraints, it is assumed in the Appendix that all the stronger O-H… bending constraints are intact, but that only half of the weaker C-H… bending constraints are intact (those associated with one flap only). This is a ***very striking result***, as it provides a natural mechanical explanation for the fact that trehalose is the preferred material for bioprotection against severe stress. This particular large-scale (very small residual or Kauzmann entropy) configuration is the one that is realized on time scales of s at 396 K, so it is easily seen that at lower temperatures it could successfully provide protection for very long times.

The dynamical behavior of trehalose has also been studied by MDS on a time scale of 10 ps, which showed the expected strong H bonding [17], and fluctuations of the sugar flaps around the dihedral angles shown in Fig.2. Simple sugars such as mannose, sucrose, fruchtose and glucose (differing chiefly in conformations) [18] are modestly overrconstrained glasses with strong H bonding (Table I). The bioprotective ordering of these sugars is the same as their viscosities [14]. However, the protective action of trehalose is enhanced by mixing it with 5% glycerol [17], which reduces the viscosity. We conclude that mean field theory, and macroscopic variables such as viscosity, only partially account for these small differences. Instead, optimal performance is achieved by the trehalose-glycerol mixture because it preserves ideally glassy molecular character, while benefiting from improvements in limiting the accumulation of network stress on protein (much longer than small molecular) length scales. Stress accumulation, which can lead to fracture, is a characteristic ferroelastic feature of adaptive nonlinear materials (including glasses) [19].

The ideality of protein – water interfaces is implicitly guaranteed by evolution, but the nature and origin of this ideality remains one of the central explicit problems in biology. Recently emphasis has shifted from continuum models of solvation of small molecules [20] to the glassy network nature of the first hydration shell around proteins [1,21]. X-ray and neutron scattering in $H_2O$ and $D_2O$ solutions showed that the first hydration shell around proteins has an average density ~ 10% larger than that of the bulk solvent [21,22]. MDS has shown that the strongest coupling is between the hydration shell and the bulk water, while the coupling of the hydration shell to the protein substrate is weak [23], a conclusion supported by experiments [1].

Constraint theory provides an independent test of the ideality of the first hydration shell. With maximal hydrogen bonding, O is effectively 4-fold coordinated, and H is effectively 2-fold coordinated. Thus $H_2O$ at low T becomes topologically isomorphic to $SiO_2$. This explains the similarities of the (P,T) phase diagram of ice at high P to that of $SiO_2$. Bulk $SiO_2$ is an ideal glass [8], and the $Si/SiO_2$ interface is the most perfect substrate/glass interface known (defect concentration $10^{-4}$). An Si*O* monolayer is sandwiched between an $SiO_2$ overlayer and the Si substrate, and one can now calculate the constraints by a symmetry argument [24]. The Si* atom in the SiO monolayer forms on the average two bonds with the Si substrate, and two bonds with the $SiO_2$ overlayer. This gives Si-Si*-Si and O-Si*-O bending constraints, and the angle between the normals to the intact Si-Si*-Si and O-Si*-O planes supplies one further bending constraint. There are 4/2 stretching constraints for Si*, and 2/2 stretching constraints for O*, giving a total of 6 bending and stretching constraints for the Si*O* monolayer, which forms a perfectly glassy interface. Repeating this argument for the hydrogen bonded first hydration shell O*H* between a protein substrate and an $H_2O$ overlayer, we conclude that the hydrogen bonded first hydration shell O*H* is also an ideally glassy interfacial layer. The 10% density enhancement can be regarded as a secondary effect associated with the relative dominance of hydrophilic interactions at the surface of the protein compared to hydrophobic ones in its core, where some water is still buried.

We now return to several amusing aspects of Table 1. As shown in the Appendix, the numbers of intact O-H… and C-H… bending constraints alone at $T_g$ = 396 K are 8 and 7, respectively, while the stretching plus bending constraints combined are 37. Trehalose has a very high ability to bind water for concentrations between 10 w% and 50 w% trehalose and for temperatures between 310 K and 350 K [25]. The hydration number varies as a function of temperature between 10 and 12 water molecules per trehalose molecule. Given that $T_g$ for water is low, one would expect a hydration number certainly < 15, and perhaps << 37, so the observed values of 10-12 are consistent with constraint counting. Finally, the Appendix shows that only glycerol and trehalose, the two extreme cases (lowest and highest $T_g$), behave simply with $N_c = 3N_A$. The other examples are intermediate cases, where $N_c \sim 3N_A$. This is a common situation in complex problems; theory is content with successes in the extreme cases, which bracket the intermediate cases.

A comment is appropriate here concerning some basic differences in methodology. MDS of glycerol and trehalose have utilized ionic models with short range repulsive pair interactions [17,26]. Ionic models are economical and in the early days were used for modeling vitreous silica. However, although economical for parameterizing partially ionic glass formers, long-range ionic interactions have their drawbacks compared to short-range valence force fields. In vitreous silica, for example, they do not conveniently explain why the Si angular distribution is narrow and the O one wide, which is the key to explaining why silica is a good glass former and alumina is not [8]. Subtle features of glass formation, including the low temperature immiscibility dome in sodium silicate [8], and the fact that $ZnCl_2$ is the only good extremely ionic glass former (with radial distribution functions very close to $GeSe_2$!) [27], are also inaccessible to point-charge MDS. Our broad experience with network structures has shown that point charge models often fail to account for the glass-forming tendency, while constraint counting based on short-range valence force fields has proved to be very reliable [28].

Discussion of the ideally glassy molecules, hydrogen-bonded glycerol and especially hydrogen-bonded trehalose, led us also to analyze hydrogen-bonded sorbitol and sugars, which have $T_g$ 's between those of glycerol and trehalose. If the C-H… bending constraints are neglected, these molecules are underconstrained, but if most of the C-H… bending constraints are intact, they can also be glassy or slightly underconstrained. Similarly, it has been suggested that protein functionality is controlled by only a few degrees of freedom [1]. (Proteins in their native (crystalline) states are about 10% overconstrained [3].) These similarities suggest a topological definition of the transition states measured by Fersht's mutagenesis experiments [29]: they are slightly underconstrained, with a number of "floppy" modes determined by the number of dihedral angles that preserve rigid subunits (like α helices) [30]. It seems likely that transition states correspond to the underconstrained edge of the reversibility-non-aging window of network glasses [4]: they are as soft as possible, while remaining nearly reversible. These analogies, together with the successes shown in Table I, bode well for flexibility-oriented minimalist models [3] and even more simplified bead models [31]. It is possible that the energies associated with a large part of the hierarchy of H bonding energies could be explored by comparing specific heat data on glasses and crystals [26,32].

Most drugs are obtained starting from naturally synthesized materials, which one can conjecture are always slightly underconstrained, when the constraints are counted following the rules developed for proteins [26] or inorganic network glass formers [33], supplemented by optimal hydrogen bonding (appropriate to an aqueous environment). It seems possible that the most promising synthetic starting materials for new drugs could be contained in the set of slightly underconstrained organic molecules. Of course, this "transition state" set is still large, but its identification is quite simple (for instance, in carbohydrates [34]), and combined with other criteria (such as molecular morphology, and temporal trends contained in largely proprietary commercial data bases) this criterion might yield viable road maps (for example, for polysaccharide-enzyme interactions [29,35]), similar in utility to the road maps employed in the microelectronics industry. Amusingly, enzyme loop closure relating distant sites (definitely a large-scale connectivity change, probably facilitated by a suitable floppy mode) occurs in polysaccharide-enzyme interactions [36].

The complexity of hydrogen bonding is well known, and the successes reported here are unexpected. They arise from a deliberate choice of simplest cases: singly bonded hydrocarbons, with no amide hydrogen bonds. The hydrocarbons generally have linear or nearly linear ground state geometries, whereas the amide bonds (especially with side chains) are bent by correlation energies associated with lone pairs [37]. Analogies with glassy networks also are less appropriate for amides; in fact, there appear to be no good amide molecular glass formers, apparently because bent hydrogen bond geometries always have much lower enthalpies in specific crystalline configurations than can be compensated entropically in more general glassy ones. Although direct analogies with glassy networks are inadequate for describing all aspects of protein dynamics, one can expect that some of the most relevant aspects, involving network stresses and long time scales, are still described correctly by the global topological factors discussed here. Combining global topology with a second local topological factor (the local ring statistics) was sufficient to predict the composition of window glass [8], and it is reasonable to suppose that similar combinations of local factors (accessible to first principles quantum calculations [38]) and global stress factors (inaccessible to first principles quantum calculations [38] and most MDS) may succeed in describing protein dynamics, especially near the transition state.

| Material | Formula | $T_g$ (K) | $3N_A$ | $N_c$ | $(N_c + 2N_{OH})$ | $(N_c + 2N_{OH} + 2N_{CH})$ |
|---|---|---|---|---|---|---|
| methane | $CH_4$ | - | 15 | 9 | 9 | 13 |
| amorphous Si | a-Si | - | 3 | 7 | 7 | 7 |
| p-butadiene | $(CH_2)_n$ | 180 | 9n | 9n | 9n | 11n |
| glycerol | $HCOH(H_2COH)_2$ | 189 | 42 | 31 | 37 | 42* |
| sorbitol | $(HCOH)_4(H_2COH)_2$ | 273 | 78 | 55 | 67 | 75 |
| trehalose | $(C_6O_5H_{11})_2O$ | 396 | 135 | 98 | 114 | 135* |
| glucose | $C_6O_6H_{12}$ | 304 | 72 | 50 | 60 | 74* |
| fructose | $C_6O_6H_{12}$ | 290 | 72 | 50 | 60 | 74 |
| vitreous silica | g-$SiO_2$ | 1540 | 9 | 9 | 9 | 9 |
| a-orpiment | g-As$_2$Se$_3$ | 463 | 15 | 15 | 15 | 15 |
| (Ge,Se) glass | g-GeSe$_4$ | 463 | 15 | 15 | 15 | 15 |
| optim.Na silicate | Na$_2$O(SiO$_2$)$_4$ | 760 | 45 | 45 | 45 | 45 |
| window glass (SiO$_2$)$_{74}$(Na$_2$O)$_{16}$(CaO)$_{10}$ | | 750 | 870 | 870 | 870 | 870 |

Table I. Glass-forming tendencies of selected materials. The ideal glass-forming condition is $3N_A = N_c + 2N_{OH} + (2)N_{CH}$, depending on whether or not the C-H…bending constraints are intact. Note that quite different results would be obtained from a mechanical spring model with 6 (rather than 5) bending constraints/tetrahedral C; in the case of glucose, $N_c + 2N_{OH} + N_{CH}$ would be increased to $76 > 3N_A = 72$, making glucose and all sugars strongly overconstrained rather than marginally constrained, with glycerol ($N_c + 2N_{OH} + N_{CH} = 45 > 3N_A = 42$) and trehalose ($N_c + 2N_{OH} + N_{CH} = 147 > 3N_A = 135$) no longer being ideal glasses. *See appendix.

**Figures and Captions**

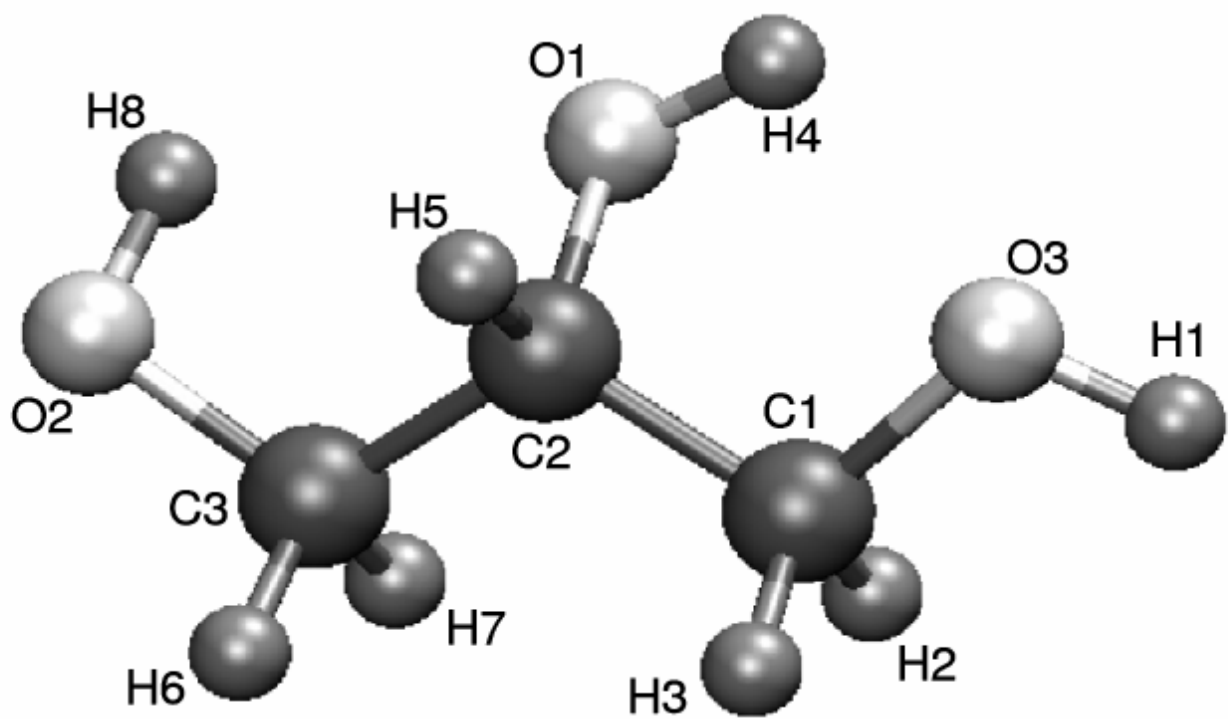

Fig. 1. The molecular structure of glycerol [17].

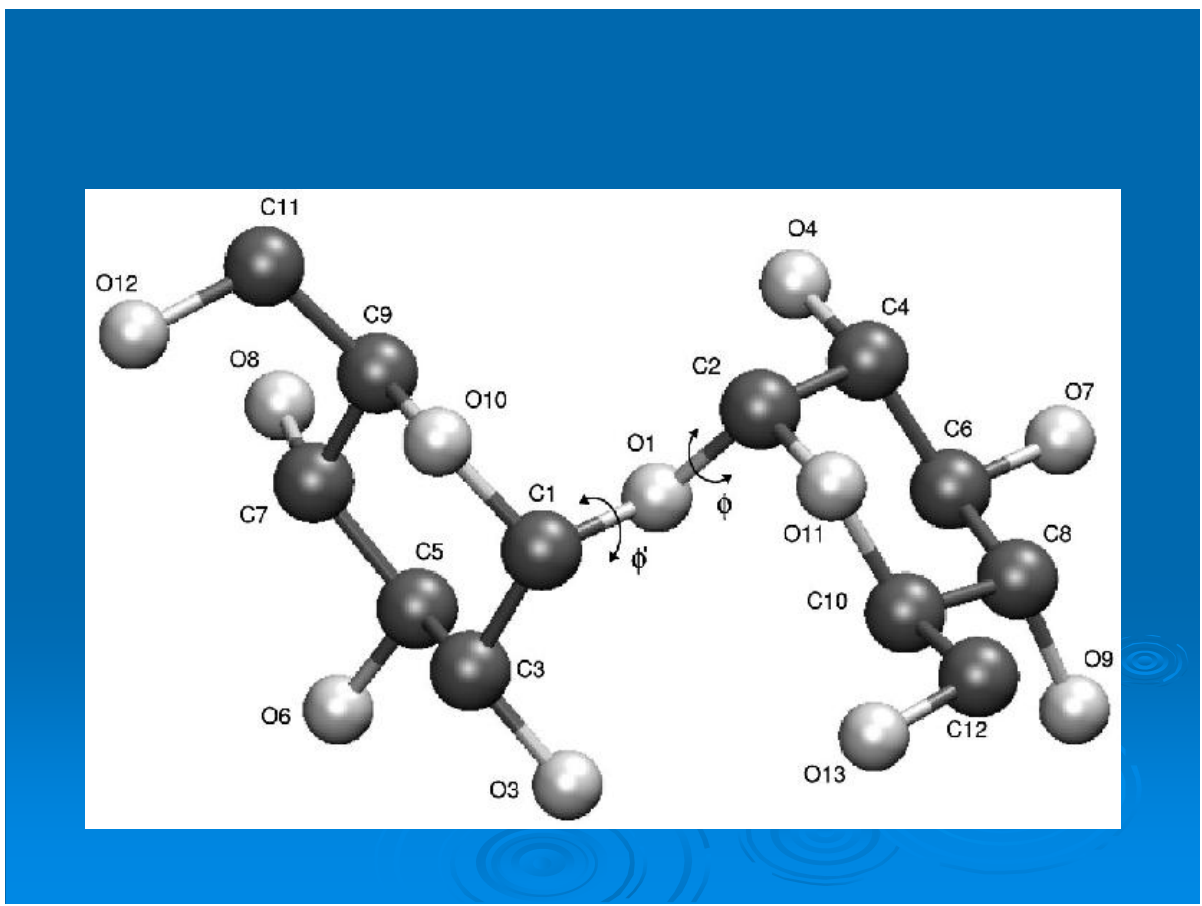

Fig. 2. The molecular structure of trehalose. The H atoms (not shown) complete the covalent intramolecular network, so that each carbon (oxygen) atom is four- (two-)fold coordinated. Intermolecular hydrogen bonds can attach to all the carbon and oxygen atoms, except for the bridging oxygen O1 [17].

**Appendix. Constraint Counting**

The rules for constraint counting in bulk network glasses are available in several papers [7-9], but they will be extended below for hydrogen bonding in glycerol, glucose and trehalose. The two-body stretching constraints are easily counted, but the three-body bending constraints are less obvious. They can be divided into intramolecular covalent and intermolecular H bonds, but one must be careful to follow consistent rules at the intra-inter-molecular interface, similar (at least in spirit) to those used at substrate-glass (Si/SiO$_2$ and protein/OH$_2$) interfaces, where only some of the bending constraints are intact. Here the C-H… bond bending constraints are weak in general, and are apparently broken in glycerol (lowest T$_g$), and are only 50% intact in trehalose (highest T$_g$).

*Glycerol* (N$_A$ = 14); Constraints **[running total]**: intramolecular stretching constraints: 13 **[13]**. intramolecular bending (2N – 3, where N is the number of single bonds): 3C, N = 4, 3(2N – 3) = 15 **[28]**, 3O, N = 2, 3(C-O-H) = 3 **[31]**. Intermolecular: stretching H…, 8 **[39]**; bending, 3(O-H…) **[42]**. (The C-H… bond has a stretching constraint, but it is too weak to have a bending constraint. Stated differently, the intact-broken gap in the constraint hierarchy at T = T$_g$ lies between the O-H… and C-H… bending constraints. Total number of constraints: **[42]** = 3 N$_A$. Ideal glass.

*Glucose* (N$_A$ = 24), carbon numbering from [18]; intramolecular stretching constraints: C$_5$O ring: 6 **[6]**; C$_5$-C$_6$:1 **[7]**; 5 C-O-H: 10 **[17]**; C$_6$–H$_2$: 2 **[19]**; intramolecular bending: 6C: 30 **[49]**; O (ring): 1 **[50]**; Intermolecular: stretching H…: 12 **[62]**; bending, (O-H…): 5 **[67]**; (C-H…) 7 **[74]**. Here most of the C-H… bending constraints are intact, as evidenced T$_g$(glucose) ~ 1.6 T$_g$(glycerol), hence glucose is slightly overconstrained, **[74]** > 3 N$_A$, and readily crystallizes in a layered structure.

*Trehalose* (N$_A$ = 45): intramolecular stretching constraints: 5CO rings: 12 **[12]**; ; C$_{9(10)}$-C$_{11(12)}$:2 **[14]**; 8 C-O-H: 16: **[30]**; C$_{11,12}$–H$_2$: 4 **[34]**; C-O$_1$-C: 2 **[36]**; intramolecular bending: 12C: 60 **[96]**; 2O (ring): 2 **[98]**; Intermolecular: stretching H…: 22: **[120]**;

bending, (O-H…): 8 **[128]**.  If one now assumes that the 7(C-H…) bond angles are satisfied for only one flap but not the other (due to the rotations around the dihedral angles), this gives 7 more constraints (not 14) for a total of **[135]** = 3 $N_A$: ideal glass. The folding of sandwich-like proteins involving interlocked pairs of neighboring β strands exhibits similar bimodal behavior [34]: half of the residues form native-like residues in the folding transition state, whereas the other half are absent from the folding state, but present in the native state.  This kind of grouping greatly reduces entropy changes and simplifies folding pathways; in the case of trehalose, it provides a natural explanation for its uniquely effective bioprotectivity.  The specific residues involved are conserved in all known sequence-diverse sandwich-like proteins; this may be the simplest example of a mechanical mechanism for evolutionary selective residue conservation.